\documentclass[sigconf,nonacm,natbib=false]{acmart}
\makeatletter
\global\@ACM@balancefalse
\makeatother
\usepackage{balance}

\setcopyright{none}
\renewcommand\footnotetextcopyrightpermission[1]{}
\usepackage[
  backend=biber,
  style=numeric,
  sorting=none,
  giveninits=true,
  maxbibnames=3,
  minbibnames=3,
  doi=false,
  isbn=false,
  url=false,
  eprint=false
]{biblatex}
\DeclareNameAlias{author}{family-given}
\DeclareNameAlias{editor}{family-given}

\renewrobustcmd*{\bibinitperiod}{}
\renewrobustcmd*{\bibinitdelim}{}
\DeclareDelimFormat{multinamedelim}{\addcomma\space}
\DeclareDelimFormat{finalnamedelim}{\addcomma\space}
\DefineBibliographyStrings{english}{andothers={et\addabbrvspace al\adddot}}
\DeclareFieldFormat[article,inproceedings,incollection,thesis,misc]{title}{#1}
\DeclareFieldFormat{pages}{#1}

\AtBeginBibliography{\sloppy}

\newbibmacro*{compact:author-title}{%
  \printnames{author}%
  \setunit{\addperiod\space}%
  \printfield{title}}

\DeclareBibliographyDriver{article}{%
  \usebibmacro{bibindex}%
  \usebibmacro{begentry}%
  \usebibmacro{compact:author-title}%
  \newunit\newblock
  \printfield{journaltitle}%
  \setunit{\addcomma\space}%
  \printfield{year}%
  \iffieldundef{volume}
    {}
    {\setunit{\addcomma\space}%
     \printfield{volume}%
     \iffieldundef{number}{}{\printtext{\mkbibparens{\printfield{number}}}}}%
  \iffieldundef{pages}{}{\setunit{\addcolon\space}\printfield{pages}}%
  \usebibmacro{finentry}}

\DeclareBibliographyDriver{inproceedings}{%
  \usebibmacro{bibindex}%
  \usebibmacro{begentry}%
  \usebibmacro{compact:author-title}%
  \newunit\newblock
  \printfield{booktitle}%
  \setunit{\addcomma\space}%
  \printfield{year}%
  \iffieldundef{pages}{}{\setunit{\addcolon\space}\printfield{pages}}%
  \usebibmacro{finentry}}

\DeclareBibliographyDriver{thesis}{%
  \usebibmacro{bibindex}%
  \usebibmacro{begentry}%
  \usebibmacro{compact:author-title}%
  \newunit\newblock
  \printtext{Master's thesis}%
  \iflistundef{institution}{}{\setunit{\addcomma\space}\printlist{institution}}%
  \setunit{\addcomma\space}%
  \printfield{year}%
  \usebibmacro{finentry}}

\DeclareBibliographyDriver{misc}{%
  \usebibmacro{bibindex}%
  \usebibmacro{begentry}%
  \usebibmacro{compact:author-title}%
  \iffieldundef{year}{}{\setunit{\addcomma\space}\printfield{year}}%
  \usebibmacro{finentry}}

\title{Research on Intellectual Property Resource Profile and Evolution Law}

\author{Yuhui Wang}
\affiliation{%
  \institution{Beijing Key Laboratory of Intelligent Communication Software and Multimedia, School of Computer, Beijing University of Posts and Telecommunications (National Demonstration Software School)}
  \city{Beijing}
  \country{China}}

\author{Yingxia Shao}
\authornote{Corresponding author.}
\affiliation{%
  \institution{Beijing Key Laboratory of Intelligent Communication Software and Multimedia, School of Computer, Beijing University of Posts and Telecommunications (National Demonstration Software School)}
  \city{Beijing}
  \country{China}}
\email{shaoyx@bupt.edu.cn}

\author{Ang Li}
\affiliation{%
  \institution{Beijing Key Laboratory of Intelligent Communication Software and Multimedia, School of Computer, Beijing University of Posts and Telecommunications (National Demonstration Software School)}
  \city{Beijing}
  \country{China}}

\begin{abstract}
In the era of big data, intellectual property-oriented scientific and technological resources show the trend of large data scale, high information density, and low value density, which brings severe challenges to the effective use of intellectual property resources, and the demand for mining hidden information in intellectual property is increasing. This makes intellectual property-oriented science and technology resource portraits and analysis of evolution become a current research hotspot. This paper sorts out the construction method of intellectual property resource portraits and its preliminary work, including property entity extraction and entity completion, from the aspects of algorithm classification and general process, and identifies directions for improving future methods.
\end{abstract}

\keywords{intellectual property, resource profile, named entity recognition, evolution analysis, deep learning}

\begin{document}
\maketitle

\section{Introduction}

As the most important information carrier and knowledge source of research results and technological innovation, patents are the main object of intellectual property analysis. The research on intellectual property in this paper also focuses on patents. With the rapid development of science and technology and the acceleration of technological iteration, the number of patents has exploded. Analysis and mining of intellectual property resources mainly based on patents can extract technology concepts, technology application fields, and other information from a large amount of patent data, and then reveal the development status and trends of technology. This helps enterprises identify technology opportunities~\cite{ou2017mainstream}, seize market opportunities~\cite{su2020technologyOpportunity}, adjust claims to improve authorization opportunities~\cite{krestel2021patentSurveyA}, and enhance their core competitiveness. Interpretable machine-learning models also provide a way to connect extracted evidence with intelligent managerial decisions~\cite{li2019interpretableDecision}.

Patent literature requires a strong professional background to understand, and its analysis mostly relies on patent analysts~\cite{li2018patentTextReview}. With the rapid increase in the number of patents, interdisciplinary technologies continue to emerge, and it is difficult to understand technological development quickly and comprehensively through manual analysis alone. Patents contain a large number of specialized words and technical terms, characterized by precise language, complex semantic information, and high information density, which challenges the accurate extraction of key information. At the same time, there are complex and rich connections among intellectual property entities such as technical concepts, applicants, and involved fields. Changes in these relationships can reflect fine-grained changes and development in intellectual property. When traditional patent analysis and data-mining methods analyze important information such as technical concepts and research topics, there is serious loss and fragmentation of semantic information~\cite{walter2022digitalization}. These methods also make insufficient use of relationships among intellectual property entities, making it difficult to capture the development and change of intellectual property in subdivided fields.

With the development of artificial intelligence and big-data technology, data profiling~\cite{liu2021dataProfile} has been used increasingly widely. Data portraits can comprehensively use extracted entity and relationship information in metadata to obtain highly refined feature representations and use these features to mine data patterns. Constructing intellectual property resource portraits transforms unstructured patent text into structured expressions such as entity-relationship records that are readily understood, effectively organizes high-density technical information in patents, and enhances deep semantic relationships among patents.

This paper introduces the main technologies used in the construction of intellectual property resource portraits and existing methods for analyzing technology evolution. It summarizes entity extraction, entity completion, portrait construction, and evolution analysis, and discusses directions for further improvement.

\section{Research on Intellectual Property Resource Profiles}

\subsection{Entity extraction method}

Named entity recognition identifies spans in text and assigns them to predefined semantic types~\cite{chen2020nerReview}. Patent-oriented recognition is more difficult than general-domain recognition because entity boundaries are long and variable, terminology is dense, and the same technical expression can have different meanings across contexts. Although state-estimation research on coupled complex networks is not itself a patent-mining method, its treatment of uncertainty and dependencies provides useful methodological context for robust modeling~\cite{li2017varianceState}. Graphical dependency modeling has also been used to combine learned and explicit correlations in semi-supervised fake-news detection~\cite{dong2023deepMRF}. Reviews of technical-term recognition further show that patent analysis must combine lexical, contextual, and domain information~\cite{hu2021termRecognition}.

Early intellectual property extraction often used subject-action-object structures and manually designed rules. Such methods have been applied to Chinese patent entity-relation extraction~\cite{zhang2019saoExtraction}, technology-theme evolution~\cite{ren2021saoEvolution}, automatic construction of technical-efficacy maps~\cite{zhao2018efficacyMap}, and the extraction of advantages and drawbacks from product descriptions~\cite{chiarello2017productDescription}. Named entity recognition and association rules have also been combined for traditional Chinese medicine patent mining~\cite{chen2020tcmPatentNER}.

Traditional statistical learning and neural sequence labeling represent two major lines of entity extraction. The first combines dependency analysis, dictionaries, rules, and statistical classifiers; the second learns representations and label dependencies from data. Rule-enhanced systems can be computationally efficient, but their performance degrades on long sentences and multiword technical entities. Work on iterative learning and stability analysis illustrates the broader need to control error propagation in sequential models~\cite{meng2010iterativeLearning}. Social-media NER demonstrates how noisy domain language requires adaptable features~\cite{ritter2011tweetNER}, while representation learning in adjacent vision and heterogeneous-network tasks shows the value of learning latent features rather than relying only on manual descriptors~\cite{fang2020cycleGAN,shi2021collaborativeFiltering}. Bi-projection fusion in omnidirectional image super-resolution provides a further example of reconciling complementary views~\cite{wang2024omnidirectionalSR}. Surveys of deep learning for patent analysis document the resulting shift toward neural representations~\cite{krestel2021patentSurveyB}.

Modern patent NER commonly contains a word-embedding layer, a feature-extraction layer, and a sequence-prediction layer. Conditional random fields are frequently used at the final layer to learn dependencies between adjacent labels. A sequence-to-sequence model with sentence classification has been used to improve named entity recognition~\cite{wang2019scNER}, and bidirectional LSTM models have been evaluated on biomedical and patent data~\cite{saad2020biLSTMPatentNER}. Gated relation networks enhance convolutional NER by explicitly modeling token relations~\cite{chen2019grn}, while contextualized embeddings improve chemical named entity recognition in patents~\cite{zhai2019chemicalNER}. Knowledge-enhanced modeling of entities and their relationships has likewise improved sarcasm detection, illustrating the value of external relational evidence for difficult language phenomena~\cite{wang2025sarcasm}. For Chinese text, lattice LSTM combines character and lexicon information~\cite{zhang2018latticeLSTM}; HMM-based bidirectional matching dynamically updates word segmentation~\cite{yan2021hmmBimm}. Sequential modeling in other high-dimensional domains also highlights the importance of representing changing states rather than treating observations independently~\cite{zhao2018p2pLending}. BiLSTM-CRF models support keyphrase extraction from scholarly documents~\cite{alzaidy2019keyphrase}, and gated self-attention strengthens LSTM-CRF Chinese NER~\cite{jin2019gatedNER}.

To accelerate training and capture local context, iterated dilated convolutional neural networks replace recurrent LSTM computation with parallel convolutions. In recent years, Transformer-based methods have been widely used in natural language processing and have achieved excellent results. Improved models based on flat-lattice Transformer representations~\cite{li2020flat} and Transformer-XL~\cite{dai2019transformerXL} are therefore increasingly used in named entity extraction. Fine-grained Chinese NER based on RoBERTa-WWM-BiLSTM-CRF provides another example of combining pretrained contextual representations with structured sequence prediction~\cite{yin2021robertaNER}.

\subsection{Entity completion method}

After entity extraction from patent text, massive knowledge consisting of a large number of entities and relationships can be obtained. However, because some key entity categories are missing, the completeness of this knowledge is low. Translation-based embedding models provide a standard basis for diagnosing and completing missing links~\cite{bordes2013transE}. Teacher-student distillation for graphs with incomplete features and structure offers a closely related mechanism for restoring missing graph information~\cite{huo2023t2gnn}. Self-supervised reciprocally contrastive learning for heterogeneous graphs offers another way to exploit attribute and topology views when annotations are scarce~\cite{huo2022heterogeneousContrastive}. For example, in patent abstracts, more than half of the patents may not clearly indicate the field covered. This requires predicting missing entities and relationships in the graph to complete graph knowledge~\cite{wang2014transH}.

There are mainly two types of entity completion methods. The first uses the graph structure of the existing graph to generate feature representations of triples~\cite{lin2015transR,ji2015transD}. When knowledge is distributed across graphs, federated cross-graph node classification can exploit shared patterns without centralizing all graph data~\cite{guan2021federatedGNN}. Reinforcement active client selection further addresses client contribution and heterogeneity in federated graph learning~\cite{wang2025activeClientSelection}. The completion method combines a given entity-relation pair with possible entities to form candidate triples, calculates the score of each triple, and obtains a completed entity according to the score. Si Jiaqi~\cite{siTextEnhancedKGC} proposed text-enhanced knowledge graph completion that introduces the source text of entity-relation triples, represents triples and source texts separately, fuses the two kinds of features, and uses the fused representation to predict entity-relation triples.

The second category assigns existing attributes to entities with missing attributes through classification, based on the assumption that entities with missing attributes and similar entities with existing attributes share the same concept. For example, when predicting tail entities from head entities and relations, the tail entities can be determined by classifying the given head-entity relations into a set of head-entity relations that share the same tail entity. This method is suitable for nodes containing textual information. She Qixing et al.~\cite{sheBayesianAttributes} used a probabilistic and statistical model based on a Bayesian network and dependencies between hypernym concepts and attributes to recommend known attributes to entities lacking attributes. Yang Yifan et al.~\cite{yangAliasCompletion} used attributes in introductory text to train a classification model that determines whether an extracted named entity is an alias, thereby completing character aliases. Pan Luming~\cite{panMultilabelCompletion} used multi-label text classification to predict missing entity-type information.

\subsection{Portrait of intellectual property resources}

In recent years, more scholars have paid attention to the importance of scientific and technological resource information~\cite{ouyang2022resourceRetrieval}. Scientific and technological information oriented semantic- and media-adversarial cross-media retrieval explicitly aligns heterogeneous resource modalities~\cite{li2022crossMediaRetrieval}. However, compared with common user profiles~\cite{graells2016dataPortraits} and scholar profiles~\cite{liangScholarProfiles}, scientific and technological resource profiles face more challenges. Multi-view scholar clustering with dynamic interest tracking further shows how scholar portraits can represent both multiple research views and temporal changes in interest~\cite{li2023scholarClustering}. On the one hand, user portraits and scholar portraits can be constructed using public datasets. In a social-media analysis scenario, Liang et al.~\cite{graells2016dataPortraits} used Twitter data to construct user portraits, and Tang et al.~\cite{tang2008arnetMiner} established the AMiner data platform using data about researchers, scientific literature, and academic activities. It provides a data basis for studying scholar portraits. In contrast, scientific and technological resource portraits lack public datasets, and their data are distributed widely on the Internet, requiring automated acquisition and portrait construction~\cite{yang2015ontologyRetrieval}. On the other hand, constructing such profiles requires accurate identification of valid information in the acquired resources~\cite{kou2016socialSearch}; the limited accuracy of existing user- and scholar-profile methods makes this task more difficult.

Existing construction methods can be divided into ontology-based, topic-based, profile-matrix-based, and semantic-mining-based methods. Euro-CRIS uses the CERIF unified description model for national research information~\cite{li2022deepNERSurvey,kremenjas2020cerif}. Knowledge graphs built from scientific and technological big data provide an entity-relation representation for resource portraits~\cite{wang2019scitechKG}. For distributed information networks, FedSIN uses federated self-adaptive learning to obtain network representations without centralizing all data~\cite{li2026fedSIN}. Domain ontologies support scientific and technological resource clustering~\cite{ge2020resourceClustering}, while topic modeling supports profile generation without predefined labels~\cite{zha2019dataless}. Two-dimensional patent-portfolio profiling combines technological content and portfolio structure~\cite{kuan2018portfolioProfiling}. Multi-label classification with dynamic semantic representations can assign several resource attributes simultaneously~\cite{wang2020multilabel}. For incomplete multi-view multi-label resources, view-category interactive sharing can jointly exploit complementary view and category information while generating missing views~\cite{ou2024viewCategoryTransformer}. Semantic-similarity attention combined with hypergraph convolution captures high-order relations among scientific publications~\cite{li2026hypergraphPublication}. Heterogeneous graph attention also supports semi-supervised classification when scientific resource descriptions are short and sparsely labeled~\cite{hu2019heterogeneousGAT}, and distributional semantics supports problem-domain ontology mining~\cite{kozerenko2019ontologyMining}.

Application platforms then integrate these components into end-to-end services. An agricultural knowledge service platform illustrates how portrait-oriented resources can support scientific and technological innovation~\cite{qin2020agriculturalPlatform}. Retrieval-oriented masked-autoencoder pretraining can strengthen the semantic representations used for resource search~\cite{xiao2022retroMAE}. Classical K-nearest-neighbor improvements~\cite{sun2009knn} and deep low-rank multi-view clustering~\cite{xue2019lowRankClustering} provide alternative classification and clustering mechanisms. Federated supervised cross-modal retrieval extends this pipeline to distributed multimodal collections while limiting direct data exchange~\cite{li2024federatedCrossModal}. Communication-efficient federated resource services can additionally combine dynamic client selection with adaptive gradient compression~\cite{pan2025rfcsc}. Together, these methods transform heterogeneous patent resources into structured, searchable, and analyzable resource portraits.

\section{Research on the Evolution Law of Intellectual Property}

The evolution law of intellectual property can be studied from changes in technical hotspots, applicant cooperation, and technology paths. Bibliometric visualization is a common starting point. CiteSpace organizes citation and keyword relationships as a knowledge map~\cite{chen2015citespace}; it has been used to identify domestic financial-technology hotspots and frontiers~\cite{huang2020fintech} and to analyze the literature on green buildings over time~\cite{shi2019greenBuilding}. Self-supervised graph co-training in sequential settings illustrates how mutually reinforcing graph views can stabilize temporal representations~\cite{xia2021graphCoTraining}. Research on dynamic state estimation, including complex-network estimation and multi-sensor multi-target tracking with registration errors, provides adjacent methodological perspectives on dynamic modeling under uncertainty~\cite{li2017recursiveState,li2013gmphdRegistration}, although patent-evolution analysis relies primarily on bibliometric and semantic networks.

Cooperation-network analysis represents applicants, inventors, or institutions as nodes and co-application or co-invention relationships as edges. Dynamic analysis of cross-regional knowledge flow and innovation cooperation reveals how regional links change~\cite{chen2019knowledgeFlow}. Similar methods have been applied to industry-university-research cooperation in the ICT industry~\cite{chen2019ictCooperation}, nanotechnology patent cooperation~\cite{liu2014nanoNetwork}, and waste-to-energy patent cooperation~\cite{ye2017wasteEnergy}. Deep modularity-based community detection can identify cohesive collaboration groups directly from network structure~\cite{yang2016modularityCommunity}. These studies generally compare network density, centrality, component structure, and the roles of influential applicants across time slices.

Technology paths can also be recovered from citation and semantic relations. A multi-main-path method avoids reducing an evolving field to a single dominant route~\cite{chen2015multiMainPath}. Filter-enhanced MLP models provide a lightweight way to encode ordered behavioral signals~\cite{zhou2022filterMLP}. Tucker-decomposition-based dataset distillation offers a complementary way to condense sequential interaction data while retaining temporal structure~\cite{zhang2025td3}. Subject knowledge networks integrate topic information with structural relations to model the mechanisms of knowledge evolution~\cite{guan2018subjectNetwork}. For industrial clusters, patent technology mining and evolution analysis based on topic models can identify emerging and declining themes~\cite{yan2019clusterEvolution}. Keyword analysis constructs clusters from co-occurrence matrices and displays trends in the influence of cluster members through the distribution of patents. Tan Tingting~\cite{tan2020patentEvolution} and Wei et al.~\cite{wei2020lda3dPrinting} used latent Dirichlet allocation to cluster patent topics and determine evolution paths according to similarities among topic words.

\section{Conclusion}

The resource portrait method for intellectual property mainly extracts technical keywords from patent text through entity extraction and then converts unstructured patent text into a structured representation of intellectual property entities after entity completion. The evolution law of intellectual property is mainly analyzed by methods such as cooperation-network analysis and keyword analysis to mine the evolution of hotspots and cooperation. The evolution of technology development and technology paths is mostly analyzed by methods such as topic clustering.

At present, patent-mining technologies support different fields of patent analysis, but these technologies also have defects that still need to be improved. More in-depth research can focus on the following aspects:

\begin{enumerate}
  \item In intellectual property evolution analysis, popularity trends often consider only co-occurrence frequency. Indicators such as time decay, applicant influence, and patent influence can be introduced so that popularity more comprehensively reflects changes in patent influence.
  \item In constructing intellectual property knowledge graphs, the types of entity relationships can be enriched, deep-learning-based entity fusion can be introduced, and entity recognition can be improved. Patent entity-relation triples can then be constructed more accurately, thereby improving the accuracy of intellectual property knowledge graphs and resource portraits.
\end{enumerate}

\section{Acknowledgements}

This work was supported by the National Key R\&D Program of China (2018YFB1402600) and the National Natural Science Foundation of China (61772083).

\balance
\printbibliography[title={References}]

\end{document}